\documentstyle[vsolj01,graphicx,natbib]{article}

\RequirePackage[T1]{fontenc}

\def\cite{\citealt}

\def\commenta{$^*$}
\def\commentb{$^\dagger$}
\def\commentc{$^\ddagger$}
\def\commentd{$^\S$}
\def\commente{$^\|$}
\def\commentf{$^\#$}

\begin{document}

\title{ASASSN-19ax: SU UMa-type dwarf nova with a long superhump}
\vskip -2mm
\title{period and post-superoutburst rebrightenings}

\author{Taichi Kato$^1$, Hiroshi Itoh$^2$, Seiichiro Kiyota$^3$,
       Tonny Vanmunster$^4$, Tam\'as Tordai$^5$,}
\author{Yusuke Tampo$^1$, Naoto Kojiguchi$^1$, Masaaki Shibata$^1$,
       Junpei Ito$^1$, Roger D. Pickard$^6$, Sjoerd Dufoer$^7$}
\author{$^1$ Department of Astronomy, Kyoto University,
       Sakyo-ku, Kyoto 606-8502, Japan}
\email{tkato@kusastro.kyoto-u.ac.jp}
\author{$^2$ Variable Star Observers League in Japan (VSOLJ),
     1001-105 Nishiterakata, Hachioji, Tokyo 192-0153, Japan}
\author{$^3$ VSOLJ, 7-1 Kitahatsutomi, Kamagaya, Chiba 273-0126, Japan}
\author{$^4$ Center for Backyard Astrophysics Belgium, Walhostraat 1A,
     B-3401 Landen, Belgium}
\author{$^5$ Polaris Observatory, Hungarian Astronomical Association,
     Laborc utca 2/c, 1037 Budapest, Hungary}
\author{$^6$ 3 The Birches, Shobdon, Leominster, Herefordshire,
     HR6 9NG, UK}
\author{$^7$ 3 Vereniging Voor Sterrenkunde (VVS), Oostmeers 122 C,
     8000 Brugge, Belgium}

\begin{abstract}
We observed ASASSN-19ax during the long outburst in
2021 September--October.  The object has been confirmed to
be an SU UMa-type dwarf nova with a superhump period
of 0.1000--0.1001~d.  This object showed two
post-superoutburst rebrightenings both in the 2019
and 2021 superoutbursts.  These observations have
established that ASASSN-19ax belongs to a group of
long-period SU UMa-type dwarf novae which show
multiple rebrightenings.
This phenomenon probably arises from premature quenching
of the superoutburst due to the weak 3:1 resonance near
the stability border of the resonance, resulting in
a considerable amount of disk mass after the superoutburst.
We noted that ASASSN-19ax is very similar to
QZ Ser, an SU UMa-type dwarf nova with an orbital
period of 0.08316~d and an anomalously hot, 
bright secondary star, in that both objects showed multiple
post-superoutburst rebrightenings at least once and that
they are bright in quiescence.
We expect that the core of the secondary in ASASSN-19ax
may be evolved as in QZ Ser.
\end{abstract}

\section{Introduction}

   ASASSN-19ax is a dwarf nova discovered by the All-Sky
Automated Survey for Supernovae (ASAS-SN, \cite{ASASSN})
at $g$=14.84 on 2019 January 12.\footnote{
  $<$http://www.astronomy.ohio-state.edu/~assassin/transients.html$>$.
}
The ASAS-SN team described this object as ``red star outburst,
matches to PS1 G=16.1, previous outburst in CRTS''.
Using ASAS-SN Sky Patrol data (\cite{ASASSN}, \cite{koc17ASASSNLC}),\footnote{
  $<$https://asas-sn.osu.edu/$>$.
}
T.K. noticed that the 2019 January outburst already
started on 2018 December 31 (peaking at $g$=12.9)
and that this outburst looked like a superoutburst
with a dip and rebrightening (vsnet-alert 22936).\footnote{
  $<$http://ooruri.kusastro.kyoto-u.ac.jp/mailarchive/vsnet-alert/22936$>$.
}
Despite the red color, this object was suspected to be
an SU UMa-type dwarf nova (vsnet-alert 22936).
Although time-resolved observations
started on 2019 January 15, the object was already
fading and no superhump-like signal was recorded.

   Another outburst was detected by Eddy Muyllaert
on 2019 September 5 (originally reported in cvnet-outburst,
and cited in vsnet-alert 23543).\footnote{
   $<$http://ooruri.kusastro.kyoto-u.ac.jp/mailarchive/vsnet-alert/23543$>$.
}
Although time-resolved observations were initiated,
this outburst faded rapidly within 5~d of
the initial detection.  No periodic modulations were
detected and this outburst should have been
a normal outburst.

\begin{figure*}
  \begin{center}
    \includegraphics[width=16cm]{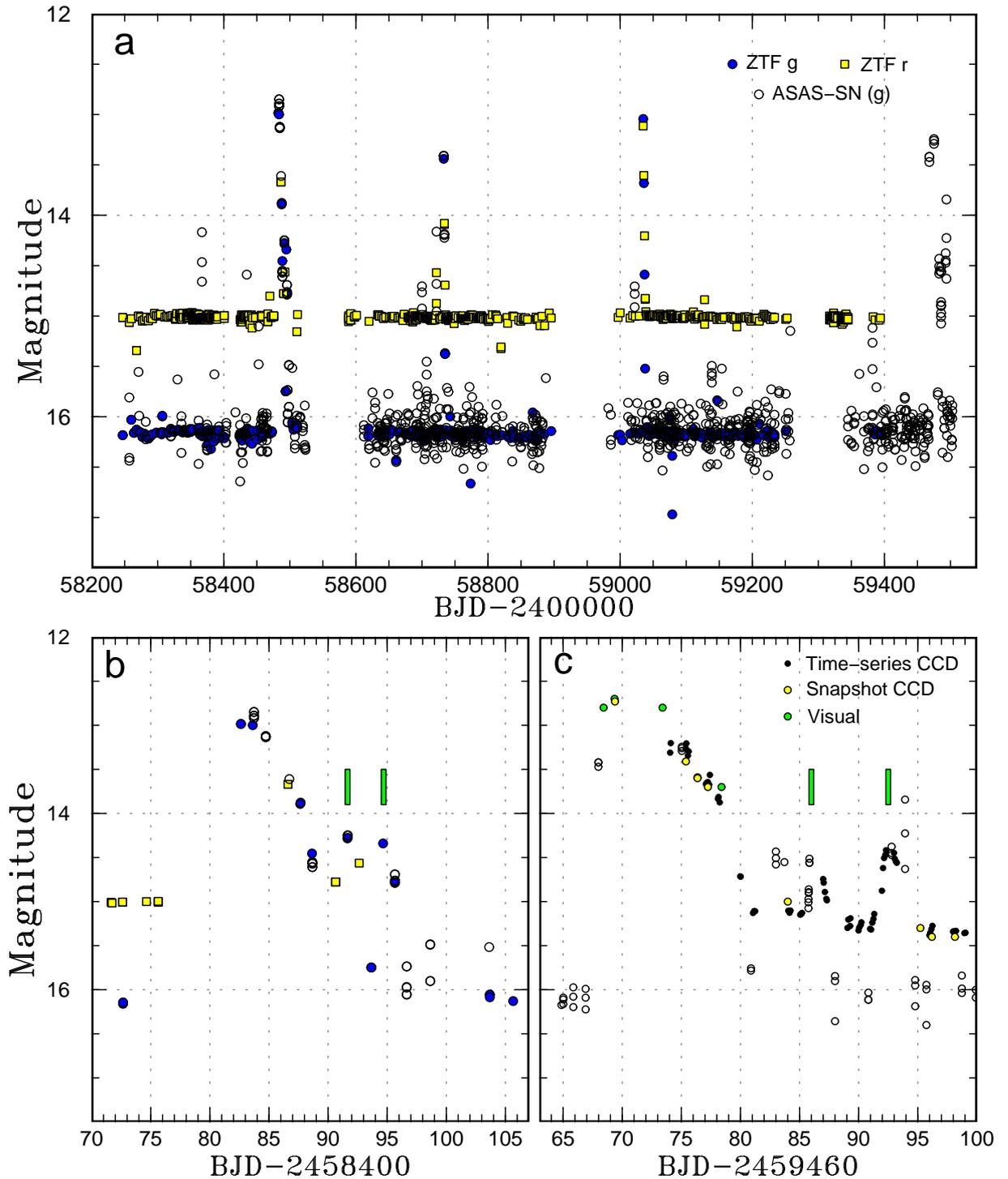}
  \end{center}
  \caption{Light curve of ASASSN-19ax.
    \textbf{a}: Long-term light curve based on ASAS-SN and
    ZTF data.  The first and last outbursts
    were superoutburst and two outbursts between them were
    normal outbursts.
    \textbf{b}: Superoutburst in 2019 January.
    Vertical green ticks represent rebrightenings.
    \textbf{c}: Superoutburst in 2021 September.
    Time-series CCD observations (this work, binned to 0.1~d),
    snapshot CCD and visual observations from VSNET and
    VSOLJ were plotted.
    Vertical green ticks represent rebrightenings.
    }
  \label{fig:lcall}
\end{figure*}

   Later on, we used Public Data Release of
the Zwicky Transient Facility \citep{ZTF}
observations\footnote{
   The ZTF data can be obtained from IRSA
$<$https://irsa.ipac.caltech.edu/Missions/ztf.html$>$
using the interface
$<$https://irsa.ipac.caltech.edu/docs/program\_interface/ztf\_api.html$>$
or using a wrapper of the above IRSA API
$<$https://github.com/MickaelRigault/ztfquery$>$.}
and confirmed that the 2019 January outburst
was followed by two rebrightenings (figure \ref{fig:lcall}b).

   Although multiple rebrightenings are usually seen in
WZ Sge-type dwarf novae \citep{kat15wzsge},
ASASSN-14ho showed four rebrightening despite its
long orbital period ($P_{\rm orb}$) of 0.24315(10)~d
\citep{gas19asassn14hov1062cyg}.  \citet{kat20asassn14ho}
suggested that ASASSN-14ho is an SU UMa star
above the period gap mimicking a WZ Sge-type dwarf nova.
There are confirmed instances of long-$P_{\rm orb}$
SU UMa-type dwarf novae with multiple rebrightenings
(e.g. ASASSN-18aan: \cite{wak21asassn18aan};
QZ Ser: VSNET unpublished, see later in this paper;
Mis V1448: N.~Kojiguchi et al. in preparation).
We therefore considered that ASASSN-19ax should be
another long-$P_{\rm orb}$ SU UMa star mimicking
a WZ Sge-type dwarf nova.

   On 2021 September 10, Eddy Muyllaert detected
a bright ($m_v$=12.8) outburst of this object.
Considering the background stated above, we launched
a campaign via VSNET Collaboration \citep{VSNET}
in vsnet-alert 26254.\footnote{
  $<$http://ooruri.kusastro.kyoto-u.ac.jp/mailarchive/vsnet-alert/26254$>$.
}  The log of observations (including the 2019 data)
is given in table \ref{tab:log}.

\begin{table}
\caption{Log of observations of ASASSN-19ax}\label{tab:log}
\begin{center}
\begin{tabular}{ccccccc}
\hline
Start\commenta & End\commenta & mag\commentb & error\commentc &
$N$\commentd & obs\commente & band\commentf \\
\hline
58499.2738 & 58499.3387 & 14.262 & 0.002 & 61 & Van & C \\
58500.2393 & 58500.3326 & 15.002 & 0.004 & 87 & RPc & V \\
58501.2928 & 58501.4191 & 15.276 & 0.003 & 138 & Van & C \\
58732.5664 & 58732.6240 & 13.289 & 0.002 & 83 & Trt & V \\
58734.1642 & 58734.2567 & 14.436 & 0.002 & 219 & Ioh & C \\
58736.5648 & 58736.6851 & 15.620 & 0.009 & 162 & DFS & V \\
58740.1789 & 58740.3169 & 15.440 & 0.006 & 233 & Ioh & C \\
59474.0042 & 59474.1054 & 13.192 & 0.005 & 207 & Kis & C \\
59475.3303 & 59475.6309 & 13.437 & 0.004 & 382 & Van & C \\
59477.0463 & 59477.3156 & 13.650 & 0.003 & 390 & Ioh & C \\
59477.0582 & 59477.2289 & 13.547 & 0.003 & 339 & Kis & C \\
59477.3102 & 59477.4182 & 13.572 & 0.006 & 142 & Trt & C \\
59478.0761 & 59478.2785 & 13.845 & 0.003 & 389 & Ioh & C \\
59479.9487 & 59480.0281 & 14.605 & 0.007 & 119 & Kis & C \\
59481.0249 & 59481.2709 & 15.118 & 0.004 & 383 & Ioh & C \\
59483.9989 & 59484.2493 & 14.981 & 0.004 & 412 & Kis & C \\
59484.0416 & 59484.2599 & 15.161 & 0.005 & 222 & Ioh & C \\
59485.0089 & 59485.2052 & 15.034 & 0.004 & 389 & Kis & C \\
59486.9912 & 59487.3250 & 14.920 & 0.007 & 264 & Ioh & C \\
59489.0201 & 59489.3104 & 15.266 & 0.007 & 260 & Ioh & C \\
59489.9769 & 59490.2410 & 15.289 & 0.005 & 260 & Ioh & C \\
59490.9299 & 59491.3262 & 15.260 & 0.004 & 666 & Ioh & C \\
59491.9420 & 59492.3263 & 14.547 & 0.007 & 495 & Ioh & C \\
59492.9612 & 59493.2458 & 14.535 & 0.003 & 349 & Ioh & C \\
59495.9646 & 59496.2602 & 15.331 & 0.003 & 503 & Ioh & C \\
59497.9578 & 59498.2561 & 15.342 & 0.002 & 466 & Ioh & C \\
59498.9098 & 59499.0653 & 15.363 & 0.007 & 235 & Ioh & C \\
59501.9186 & 59502.3293 & 15.294 & 0.004 & 331 & Ioh & C \\
59501.9596 & 59502.2411 & 1.662 & 0.001 & 778 & KU1 & C \\
59502.9103 & 59503.3120 & 15.315 & 0.003 & 281 & Ioh & C \\
59502.9591 & 59503.1697 & 1.637 & 0.001 & 580 & KU1 & C \\
59504.9101 & 59505.1489 & 15.388 & 0.007 & 147 & Ioh & C \\
59507.9012 & 59508.0267 & 15.468 & 0.006 & 209 & Ioh & C \\
59510.8743 & 59511.0001 & 15.430 & 0.003 & 226 & Ioh & C \\
59511.0784 & 59511.1653 & 1.572 & 0.003 & 188 & KU1 & C \\
\hline
\end{tabular}
\end{center}
\begin{center}
\begin{tabular}{p{10cm}}
\commenta JD$-$2400000.\\
\commentb Mean magnitude.\\
\commentc 1$\sigma$ of the mean magnitude.\\
\commentd Number of observations.\\
\commente Observer's code:
DFS (S. Dufoer),
Ioh (H. Itoh),
KU1 (Kyoto U., campus obs.),
Kis (S. Kiyota),
RPc (R. Pickard),
Trt (T. Tordai),
Van (T. Vanmunster) \\
\commentf Filter. ``C'' means unfiltered.\\
\end{tabular}
\end{center}
\end{table}

\section{Results and Discussion}

   The initial observations on 2021 September 16 by S.K.
already suggested the presence of superhumps
(vsnet-alert 26256).\footnote{
  $<$http://ooruri.kusastro.kyoto-u.ac.jp/mailarchive/vsnet-alert/26256$>$.
}
T.V. reported the detection of superhumps with a period
of 0.1015(24)~d (vsnet-alert 26260).\footnote{
  $<$http://ooruri.kusastro.kyoto-u.ac.jp/mailarchive/vsnet-alert/26260$>$.
}
This period was refined by using observations of
three observers to 0.10012(2)~d (vsnet-alert 26263).\footnote{
  $<$http://ooruri.kusastro.kyoto-u.ac.jp/mailarchive/vsnet-alert/26263$>$.
}

   After removing the global trend of the outburst by using
locally-weighted polynomial regression (LOWESS,
\cite{LOWESS}), the times of superhumps maxima were determined
by the template fitting method as described in \citet{Pdot}.
The resultant times of superhump maxima are given in
table \ref{tab:shmax}.  There appears to be a phase
jump between $E$=59 and $E$=70.  This epoch corresponds
to the termination of the initial superoutburst
and this phase jump was most likely caused by
transition to (traditional) late superhumps\footnote{
   Many classical examples of late superhumps
   in short-$P_{\rm orb}$ systems, however,
   turned out to be stage C superhumps as described
   in \citet{Pdot}.
}
(cf. \cite{hae79lateSH}; \cite{vog83lateSH};
\cite{vanderwoe88lateSH}) arising from the hot spot.

\begin{figure*}
  \begin{center}
    \includegraphics[width=15cm]{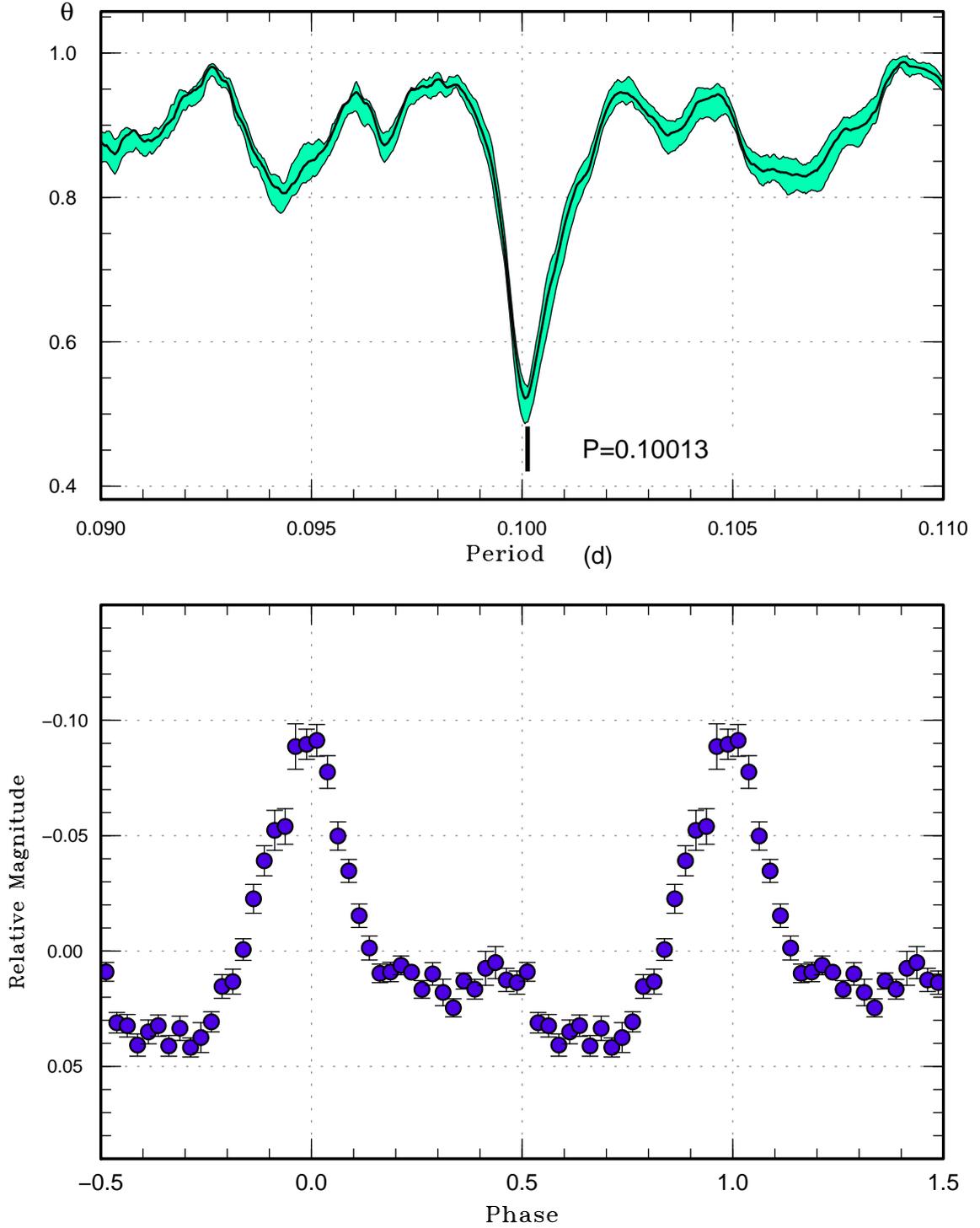}
  \end{center}
  \caption{Superhump profile of ASASSN-19ax during the 2021
    superoutburst (before BJD 2459481).
    (Upper): PDM analysis.
    We analyzed 100 samples which randomly contain 50\% of
    observations, and performed the PDM analysis for these samples.
    The bootstrap result is shown as a form of 90\% confidence intervals
    in the resultant PDM $\theta$ statistics.
    (Lower): Phase-averaged profile.
    }
  \label{fig:shpdm}
\end{figure*}

\begin{figure*}
  \begin{center}
    \includegraphics[width=15cm]{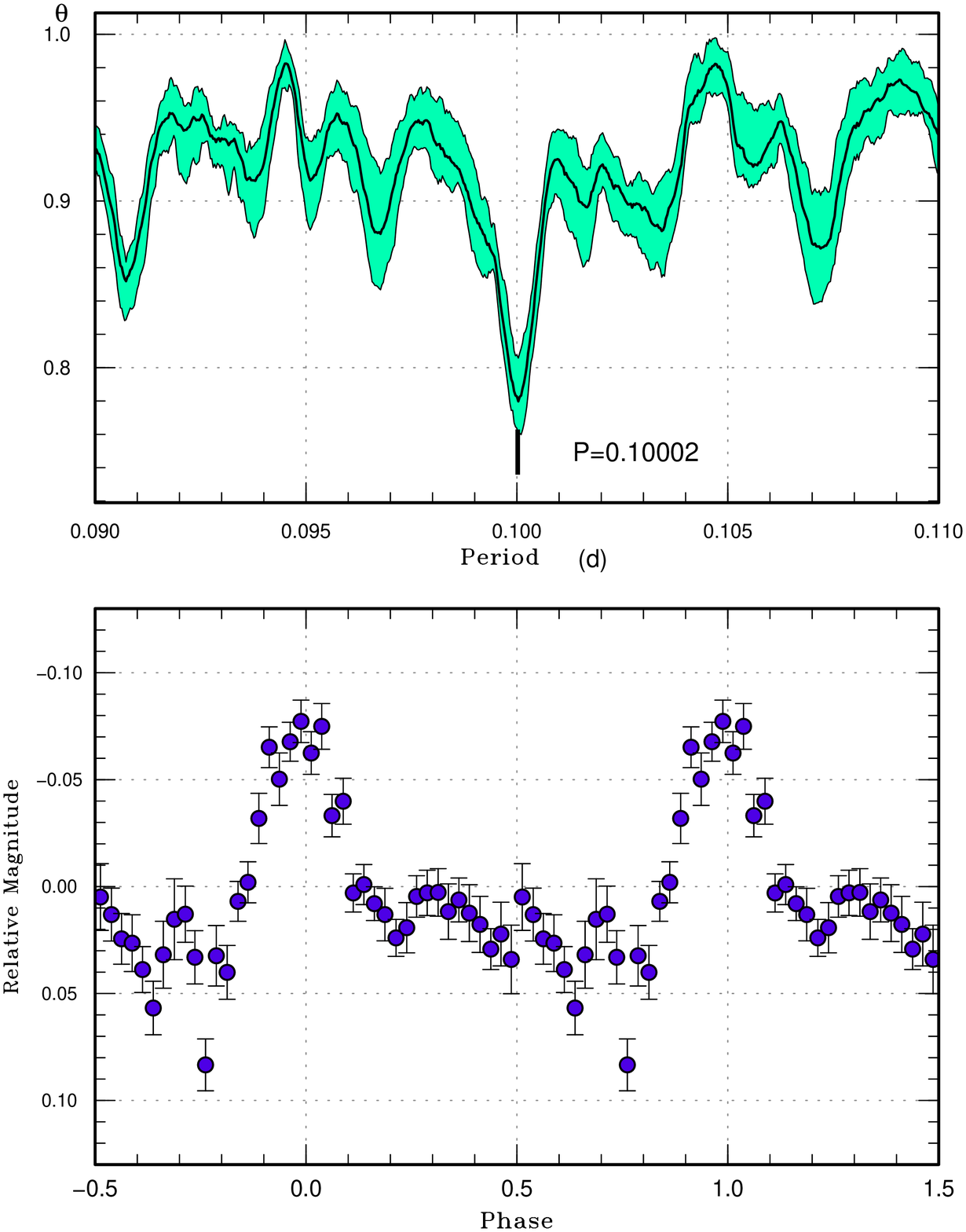}
  \end{center}
  \caption{Superhump profile of ASASSN-19ax after the 2021
    superoutburst (BJD 2459481--2459486).
    (Upper): PDM analysis.
    (Lower): Phase-averaged profile.
    }
  \label{fig:shlatepdm}
\end{figure*}

   The superhump periods were determined by
Phase Dispersion Minimization
(PDM; \cite{PDM}) method after removing the global trend
using LOWESS.  The errors were estimated by the methods
of \citet{fer89error} and \citet{Pdot2}.
The profiles and periods were almost
identical during the superoutburst (period = 0.10013(2)~d,
figure \ref{fig:shpdm}) and the post-superoutburst phase
(period = 0.10002(4)~d, figure \ref{fig:shlatepdm})
despite a phase jump between them.
Superhumps became below the detection limit after BJD 2459485.3.
Based on our observations and ASAS-SN data,
two post-superoutburst rebrightenings occurred on
BJD 2459486 and 2459492.5 (figure \ref{fig:lcall}c).
Superhumps disappeared after the first rebrightening.
No orbital signal was detected in the ZTF data
in quiescence.

\begin{table}
\caption{Superhump maxima of ASASSN-19ax}\label{tab:shmax}
\begin{center}
\begin{tabular}{rp{55pt}p{40pt}r@{.}lr}
\hline
\multicolumn{1}{c}{$E$} & \multicolumn{1}{c}{max\commenta} & \multicolumn{1}{c}{error} & \multicolumn{2}{c}{$O-C$\commentb} & \multicolumn{1}{c}{$N$\commentc} \\
\hline
0 & 59474.1147 & 0.0032 & 0&0105 & 76 \\
13 & 59475.4055 & 0.0004 & $-$0&0027 & 148 \\
14 & 59475.5158 & 0.0052 & 0&0073 & 80 \\
15 & 59475.6048 & 0.0012 & $-$0&0041 & 54 \\
30 & 59477.1103 & 0.0005 & $-$0&0031 & 300 \\
31 & 59477.2088 & 0.0005 & $-$0&0049 & 221 \\
32 & 59477.3122 & 0.0010 & $-$0&0018 & 106 \\
33 & 59477.4115 & 0.0008 & $-$0&0028 & 69 \\
40 & 59478.1158 & 0.0006 & $-$0&0007 & 145 \\
41 & 59478.2101 & 0.0016 & $-$0&0067 & 146 \\
59 & 59480.0031 & 0.0032 & $-$0&0192 & 102 \\
70 & 59481.1384 & 0.0018 & 0&0128 & 118 \\
71 & 59481.2451 & 0.0012 & 0&0191 & 154 \\
99 & 59484.0330 & 0.0019 & $-$0&0015 & 158 \\
100 & 59484.1382 & 0.0009 & 0&0033 & 200 \\
101 & 59484.2342 & 0.0016 & $-$0&0009 & 188 \\
109 & 59485.0387 & 0.0013 & 0&0011 & 130 \\
111 & 59485.2325 & 0.0086 & $-$0&0057 & 34 \\
\hline
  \multicolumn{6}{l}{\commenta BJD$-$2400000.} \\
  \multicolumn{6}{l}{\commentb Against max $= 2459474.1042 + 0.100306 E$.} \\
  \multicolumn{6}{l}{\commentc Number of points used to determine the maximum.} \\
\end{tabular}
\end{center}
\end{table}

   These observations have established that ASASSN-19ax
belongs to a group of long-$P_{\rm orb}$ SU UMa-type
dwarf novae which show multiple rebrightenings.
This phenomenon probably arises from premature quenching
of the superoutburst due to the weak 3:1 resonance near
the stability border of this resonance, resulting in
a considerable amount of disk mass after the superoutburst
\citep{kat20asassn14ho}.  The parameters for these
objects are summarized in table \ref{tab:mulreb}.

\begin{figure*}
  \begin{center}
    \includegraphics[width=15cm]{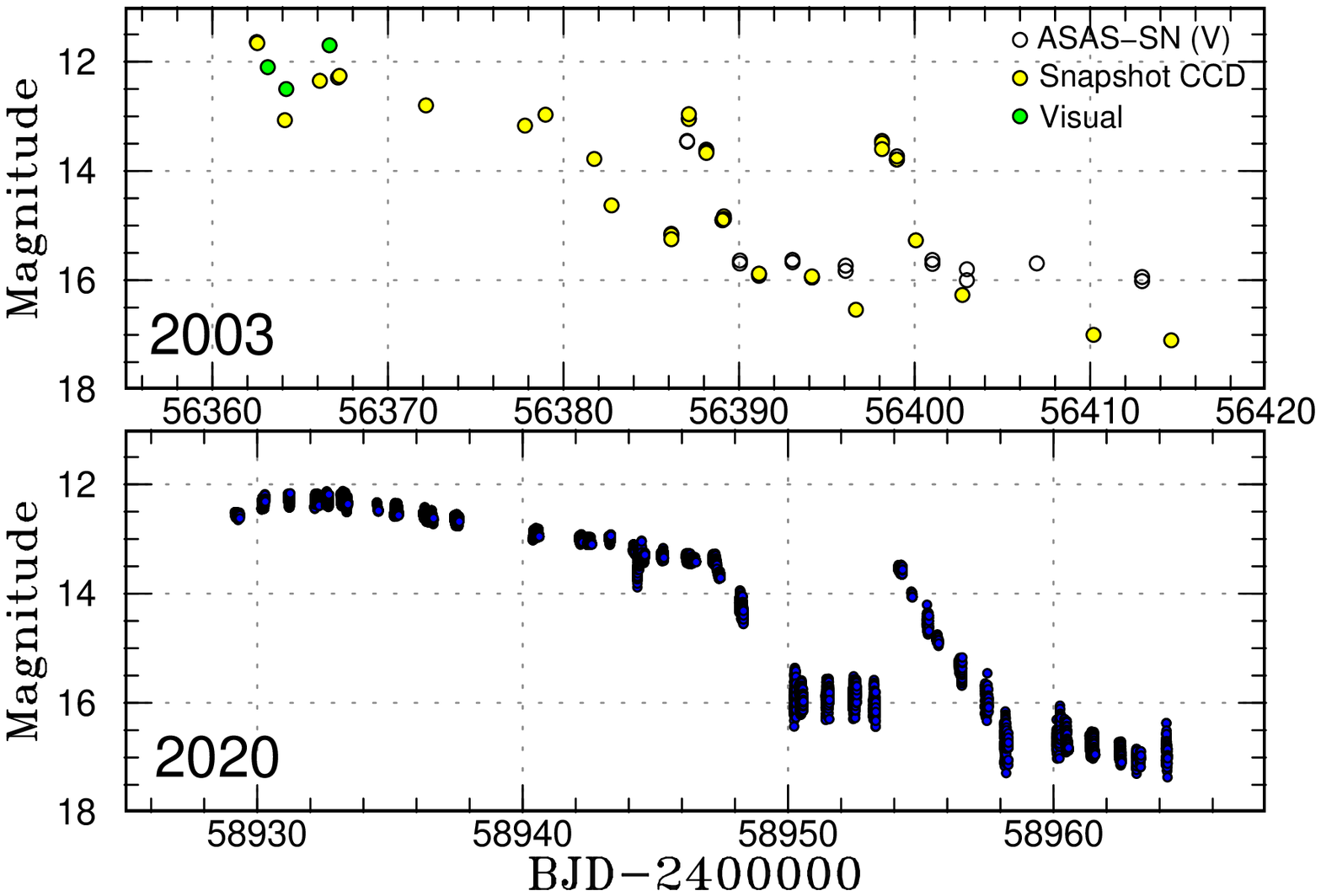}
  \end{center}
  \caption{Light curves of superoutbursts of QZ Ser.
    (Upper): Superoutburst in 2013.  ASAS-SN $V$ data,
    visual and CCD observations reported to VSNET were used.
    Two post-superoutburst rebrightenings were present.
    (Lower): Superoutburst in 2020.
    The data were from VSNET Collaboration.
    The data were binned to 0.1~d.  A post-superoutburst
    rebrightening is apparent.
    }
  \label{fig:qzser2020lc}
\end{figure*}

   The quiescent $M_V=+$7.6 \citep{GaiaEDR3} is much
brighter than $M_V=+$12 expected from the orbital period
\citep{kni06CVsecondary}.  This implies that the secondary
has an evolved core as in QZ Ser, an SU UMa-type dwarf nova
with a 2-hour orbital period and an anomalously hot, bright
secondary star \citep{tho02qzser}.  QZ Ser also showed
two post-superoutburst rebrightenings after the 2013
March-April superoutburst and one rebrightening in 2020
March-April superoutburst (\cite{kah20qzser},
although rebrightenings are not shown).
Light curves of QZ Ser are given in figure \ref{fig:qzser2020lc}
for a comparison.
The similarity between ASASSN-19ax and QZ Ser appears
to be striking.  With the bright ($r$=15.0) quiescence,
detailed spectroscopy of ASASSN-19ax is very promising.
There remains, however, a possibility that this $r$=15.0
object is not the secondary of the ASASSN-19ax binary.
The Gaia parallax of this object gives $M_V=+$5.3
for the maximum of ASASSN-19ax, which is not inconsistent
with that of a dwarf nova in outburst, and the $r$=15.0
object is unlikely a background, unrelated star.

\begin{table}
\caption{Long-$P_{\rm orb}$ (suspected) SU UMa-type dwarf novae
with multiple rebrightenings}\label{tab:mulreb}
\begin{center}
\begin{tabular}{ccccc}
\hline
Object & Orbital    & Superhump  & Number of short & References \\
       & period (d) & period (d)\commenta & rebrightenings & \\
\hline
QZ Ser & 0.083161(1) & 0.08557(13) & 1--2 & \citet{tho02qzser}, \\
       &             &             &      & vsnet-alert 15533 \\
OT J002656.6$+$284933 = & -- & 0.13225(1) & 2 & \citet{kat17j0026} \\
CSS101212:002657$+$284933 &   &  &      & \\
ASASSN-18aan & 0.149454(3) & 0.15821(4) & 2 & \citet{wak21asassn18aan} \\
Mis V1448 & --       & 0.2275(3)   & 5    & vsnet-alert 24912, N. Kojiguchi \\
       &             &             &      & et al. in preparation \\
ASASSN-14ho\commentb
       & 0.24315(10) & --          & 4    & \citet{kat20asassn14ho} \\
OGLE-BLG-DN-0174\commentb
       & --          & 0.14474(4)\commentc & 3 & \citet{mro15OGLEDNe} \\
OGLE-BLG-DN-0595\commentb
       & --          & 0.0972(1)\commentc  & 2 & \citet{mro15OGLEDNe} \\
ASASSN-19ax  & --    & 0.10013(2)  & 2    & This work \\
\hline
  \multicolumn{5}{l}{\commenta For stage B superhumps when available.} \\
  \multicolumn{5}{l}{\commentb Suspected SU UMa star.} \\
  \multicolumn{5}{l}{\commentc Requires confirmation (cf. \cite{kat17j0026}).} \\
\end{tabular}
\end{center}
\end{table}

   Although the 2021 September superoutburst was visually
detected early, time-series CCD observations started
5~d later and we probably missed the growing stage
of superhumps (stage A superhumps).  During the next
superoutburst, observations of the early stage will be
very important since the period of stage A superhumps
is vital for determining
the mass ratio \citep{kat13qfromstageA}, which might be
anomalous considering the similarity with QZ Ser.

\section*{Acknowledgments}

This work was supported by JSPS KAKENHI Grants Number 21K03616 (T.K.)
and Number 21J22351 (Y.T.).

The authors are grateful to observers of VSNET and
VSOLJ who supplied vital data.  Yasuyuki Wakamatsu
helped us in compiling data reported to VSNET.

Based on observations obtained with the Samuel Oschin 48-inch
Telescope at the Palomar Observatory as part of
the Zwicky Transient Facility project. ZTF is supported by
the National Science Foundation under Grant No. AST-1440341
and a collaboration including Caltech, IPAC, 
the Weizmann Institute for Science, the Oskar Klein Center
at Stockholm University, the University of Maryland,
the University of Washington, Deutsches Elektronen-Synchrotron
and Humboldt University, Los Alamos National Laboratories, 
the TANGO Consortium of Taiwan, the University of 
Wisconsin at Milwaukee, and Lawrence Berkeley National Laboratories.
Operations are conducted by COO, IPAC, and UW.

The ztfquery code was funded by the European Research Council
(ERC) under the European Union's Horizon 2020 research and 
innovation programme (grant agreement n$^{\circ}$759194
-- USNAC, PI: Rigault).

ASAS-SN is supported by the Gordon and Betty Moore
Foundation through grant GBMF5490 to the Ohio State
University and NSF grant AST-1515927.

\end{document}